\documentclass[api,reprint,amsmath,amssymb,superscriptaddress,showpacs,floatfix]{revtex4-1}
\usepackage{graphicx}
\usepackage{dcolumn}
\usepackage{bm}
\usepackage[colorlinks, allcolors=blue]{hyperref}
\usepackage[usenames, dvipsnames]{color}
\usepackage{setspace}
\usepackage{scalerel}
\usepackage{stackengine}
\usepackage{comment}
\usepackage{amsmath}
\providecommand{\keywords}
\stackMath
\newsavebox\tmpbox

\begin{document}

\title{Study of magnetization relaxation in molecular spin clusters using an innovative kinetic Monte Carlo method}

\author{Sumit Haldar}
\email{sumithaldar@iisc.ac.in}
\affiliation{Solid State and Structural Chemistry Unit, Indian Institute of Science, Bengaluru - 560012, India.}

\author{S. Ramasesha}
\email{Corresponding Author: ramasesh@iisc.ac.in}
\affiliation{Solid State and Structural Chemistry Unit, Indian Institute of Science, Bengaluru - 560012, India.}


\begin{abstract}
Modeling blocking temperature in molecular magnets has been a long standing problem in the field of molecular magnetism. We investigate this problem using a kinetic Monte Carlo (kMC) approach on an assembly of 100,000 short molecular magnetic chains (SMMCs), each of six identical spins with nearest neighbour anisotropic ferromagnetic exchange interactions. Each spin is also anisotropic with an uniaxial anisotropy. The site spin on these SMMCs take values $1$, $3/2$ or $2$. Using eigenstates of these SMMCs as the states of Markov chain, we carry out a kMC simulation starting with an initial state in which all SMMCs are completely spin polarized and assembled on a one-dimensional lattice so as to experience ferromagnetic spin-dipolar interaction with each other. From these simulations we obtain the relaxation time $\tau_r$ as a function of temperature and the associated blocking temperature. We study this for different exchange anisotropy, on-site anisotropy and strength of dipolar interactions. The magnetization relaxation times show non-Arrhenius behaviour for weak on-site interactions. The energy barrier to magnetization relaxation increases with increase in on-site anisotropy, exchange anisotropy and strength of spin dipolar interactions; more strongly on the last parameter. In all cases the barrier saturates at large on-site anisotropy. The barrier also increases with site spin. The large barrier observed in rare-earth single ion magnets can be attributed to large dipolar interactions due to short intermolecular distances, owing to their small size and large spin of the rare earth ion in the molecule.
\end{abstract}

\maketitle

\section{\label{sec:introduction}Introduction}
The field of molecular magnetism began with the observation of bulk magnetization in the molecular magnet by Miller $\textit{et.}$ $\textit{al.}$ in 1986 \cite{C39860001026}. Five years later, Gatteschi $\textit{et.}$ $\textit{al.}$\cite{doi:10.1021/ja00015a057}, observed magnetic polarization in the molecule $\textrm Mn_{12}Ac$ at low temperature, which heralded the field of single molecule magnets (SMMs). The discovery of SMMs raised hopes of their application in magnetic memory devices \cite{Coronado2009, Gatteschi2003, Gatteschi2003a, Palii2011, Mannini2009}. However, the low thermal barrier to magnetization relaxation, which leads to loss of magnetic memory, belied these hopes. The main focus of molecular magnetism has, therefore, been on raising the blocking temperature for magnetization relaxation by increasing magnetic anisotropy barrier. Several earlier studies focused on analyzing the effect of on-site anisotropy as well as exchange anisotropy on magnetic anisotropy barrier \cite{Rajamani2008, Haldar2017, Haldar2018, Haldar2020}.  Single chain magnets (SCMs) were subsequently synthesized, with the expectation that SCMs will have a higher blocking temperature \cite{Caneschi2001, Clerac2002, Caneschi2002, Caneschi2002, Coulon2006}. However, this has not been borne out by experiments. Recently, single rare earth ion molecular systems have been synthesized which show high blocking temperatures \cite{C0CS00185F, doi:10.1002/anie.201705426, Goodwin2017}. 

In the molecular systems, at very low-temperatures, a slow relaxation of the magnetized state occurs due to quantum resonant tunnelling which is temperature independent; at higher temperatures, the relaxation occurs due to thermally activated barrier crossing, which is assumed to follow an Arrhenius law. The temperature dependence of the relaxation times is modelled using Arrhenius expression,
\begin{eqnarray}
\label{eqn:Arrhenius}
\frac{1}{\tau_r} &=& \frac{1}{\tau_0} \textrm exp \left(-\frac{U_{B}}{K_BT}\right),
\end{eqnarray}
where $\tau_0$ is the characteristic relaxation time, $U_B$ is the thermal barrier to relaxation and $k_B$ is the Boltzmann constant and the associated blocking temperature $T_B$ is defined as $\frac{U_B}{k_B}$. Experimentally, $T_B$ is obtained from ac magnetic susceptibility measurements by identifying the peak frequency at a given temperature with $\tau^{-1}_r(T)$ and fitting the data to eqn. \ref{eqn:Arrhenius}. $T_B$ also has an operational definition; it is the temperature at which the relaxation time $\tau_r$ is 100 secs \cite{Gatteschi2006a}. It is interesting to note that the $T_B$ obtained from experiments does not correlate with the barrier height between two fully and oppositely polarized states of the SMM or SCM due to anisotropy and depends upon various scattering processes in the system. This is because the barrier crossing does not occur in a single step for activated processes and for the tunnelling process, the barrier height is largely irrelevant.

The mechanisms that contribute to magnetization relaxation can be classified into two categories (i) spin-lattice relaxation mechanisms (ii) spin-spin relaxation mechanisms. The processes involving spin-lattice interactions are the Direct, Orbach and Raman processes. As the name suggests, in the direct process, the change in magnetization of the system is followed by the creation or annihilation of a phonon. In the Orbach process, the magnetic state is excited to a higher energy vibrational state which then crosses over to different magnetized state and a lower energy vibrational state. In the Raman process, the intermediate vibrational state is a virtual state and is a hence a higher order quantum process. The inverse relaxation time in a one-phonon or direct process is linearly dependent on temperature, while the same in Raman process depends on the ninth power of temperature $(T^9)$ and the Orbach process has an exponential dependence on temperature. Spin-spin relaxation mechanisms consider both the hyperfine and dipolar interaction terms. In the present study, we considered only the role of spin-dipolar interactions. Indeed the importance of spin dipolar interactions were recognized earlier in molecular magnets \cite{Bar, Anand}. Computing magnetization relaxation times from first  principles is replete with problems such as computation of the matrix elements of the perturbation operator and computation of the phonon density of states. We do not consider the quantum resonant process, which also can lead to magnetization relaxation due to fluctuating internal magnetic field, as this is a slow and temperature independent process and dominates only at very low temperature.

In this paper, we employ kinetic Monte Carlo (kMC) simulation to estimate the blocking temperature of an assembly of 100,000 SMMCs. Each SMMC consists of six uniaxially anisotropic spins, interacting via anisotropic ferromagnetic exchange interactions. The SMMCs are arranged on a one-dimensional lattice and they interact via spin dipolar interactions. We carry out a kinetic Monte Carlo simulation of this system to obtain relaxation times for the magnetization from a fully polarized state, as a function of temperature and model parameters. From the $\tau_r(T)$ data, we obtain $T_B$ by fitting to the expression in eqn. (\ref{eqn:Arrhenius}). We have studied the dependence of $T_B$ on the strength of on-site anisotropy, magnitude of anisotropy in the exchange interactions, the strength of spin dipolar interactions which depends upon the intermolecular separations as well as the magnitude of the site spins. The paper is organized as follows: in the next section, we discuss the Hamiltonian of an SMMC and the dipolar interactions. In section III, we outline the kMC method we have employed in this study. In section IV we discuss the results for these systems as a function of model parameters and site spins. We conclude the paper with a summary and possible extensions of this work.
\section{Model Hamiltonian}
Each SMMC consists of six identical uniaxially anisotropic spins with nearest neighbour anisotropic exchange interactions. The SMMCs we have studied have site spins $1$, $3/2$ and $2$. The Hamiltonian of the system is given by,
\begin{equation}
\begin{split}
\hat{\mathcal{H}}_{SMMC} &= -J \sum_{i=1}^5 \left[\hat{s}_i^z \hat{s}_{i+1}^z+ \frac{1-\epsilon}{2} \left(\hat{s}_i^{+}\hat{s}_{i+1}^{-} + \hat{s}_i^{-}\hat{s}_{i+1}^{+} \right)\right] \\
&\quad - d \sum_{i=1}^6 \hat{s}_{i}^{z^2}.
\end{split}
\end{equation}
The anisotropic exchange interactions are restricted to the XXZ model, we have the Ising model for $\epsilon=1$ and the isotropic Heisenberg model for $\epsilon=0$. The last term represents the contribution due to the anisotropy of site spins which is assumed to be uniaxial ($d>0$), although in general the site anisotropy parameter is a tensor. The exchange interaction $J$ is taken to be ferromagnetic, hence $J>0$ and is set to unity to set the energy scale and $d$ is expressed in units of $J$.

The above Hamiltonian does not conserve $S^2$, the total spin, for nonzero $\epsilon$, it conserves $S^z$, the z-component of the total spin. We can exploit this symmetry to obtain all the eigenstates of the Hamiltonian in all the $M_s$ sectors of the full Fock space. The Fock space dimensions of SMMCs for $s=1$, $3/2$ and $2$ cases of the Hamiltonian are $729$, $4096$ and $15,625$ respectively. With full diagonalization of the SMMC Hamiltonian we have the $M_s$ and energy eigenvalues of all the eigenstates. The $M_s$ values vary between $-6s$ and $+6s$ ($s=1$, $3/2$ and $2$), all in steps of one.

\begin{figure}
    \includegraphics[width=\columnwidth]{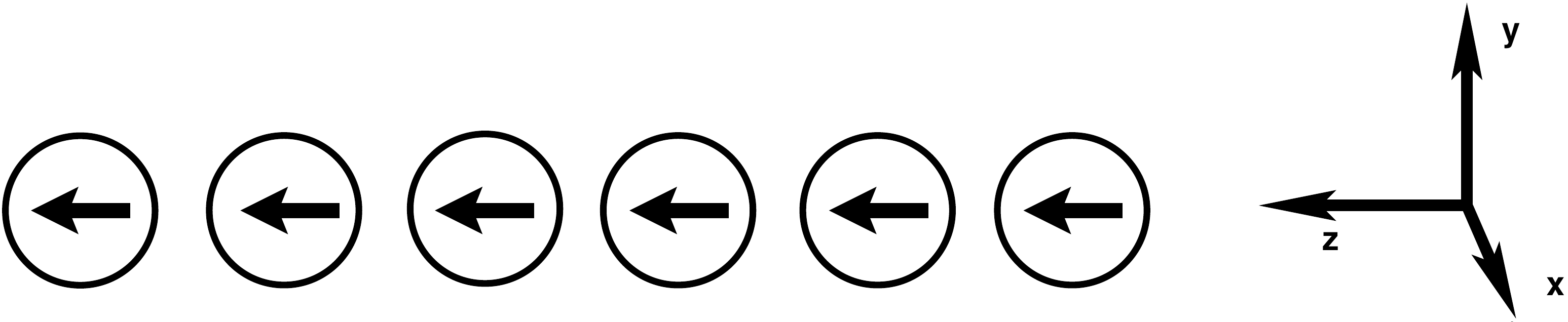}
      \caption{\label{fig:1dchain}Schematic Alignment of magnetic moments of SMMCs on the 1-d lattice in its ground state. Arrows represents the fully magnetized state of the SMMC, and the lattice is aligned along the z-axis.}
\end{figure}

In the system we study using kMC, we consider an assembly of 100,000 SMMCs arranged on a uniform one-dimensional lattice. These SMMCs interact with each other via spin dipolar interactions given by, 
\begin{eqnarray}
\label{eqn:dipolar}
\hat{\mathcal{H}}_{dip}= g^2\mu_B^2 \sum_{i>j} \frac{\vec{S}_i \cdot \vec{S}_j}{r_{ij}^3} - 3 \frac{(\vec{S}_i \cdot \vec{r}_{i})(\vec{S}_{j} \cdot \vec{r}_{j})}{r_{ij}^5},
\end{eqnarray}
where $\vec{r_i}$ are the position vectors of the center of the $i^{th}$ SMMC in the 1-d lattice and we assume the inter SMMC distance to be much larger than the inter-site distance within the SMMC. The dipolar interaction energy $\hat{\mathcal{E}}_{dip}$ in first order is given by
\begin{equation}
\begin{split}
\hat{\mathcal{E}}_{dip} &= \left\langle S_i,M_i;S_j,M_j| \hat{\mathcal{H}}_{dip} |S_i,M_i;S_j,M_j \right\rangle  \\
&\quad = g^2\mu_B^2\Biggl[-\frac{2S_i^z S_j^z}{r_{ij}^3}+  \\
&\quad \frac{3}{2}\frac{\left\langle S_i,M_i;S_j,M_j| s_i^+ s_j^-+s_i^-s_j^+|S_i,M_i;S_j,M_j \right\rangle}{r_{ij}^3}\Biggr]
\end{split}
\end{equation}
The second term goes to zero in first order and can be neglected. Hence for any geometry the contribution to $\hat{\mathcal{E}}_{dip}$ in first order is only $-\frac{2S_i^z S_j^z}{r_{ij}^3}$. If the magnetic moments of the SMMCs are initially oriented perpendicular to the direction of the 1-d lattice, the interaction between the moments will be antiferromagnetic and the ground state will be nonmagnetic. To have a fully magnetized state as the ground state, we orient the site magnetic moments along the 1-d lattice (Fig. \ref{fig:1dchain}). Assuming that the lattice constant of the 1-d lattice is unity, we can write the spin-dipolar interaction term as,
\begin{eqnarray}
\hat{\mathcal{H}}_{dip}=Cg^2\mu_B^2 \left(\sum_{i>j} (-2) \frac{S_i^z S_j^z}{|i-j|^3} \right).
\end{eqnarray}
Here, we have introduced the parameter $C$, which is used to vary the strength of intermolecular interactions which in turn depends upon intermolecular separation. The dipolar interaction energy between two fragments in eigenstates with magnetizations $M_i$ and $M_j$ and located at sites `$i$' and `$j$' in the 1-d lattice is given by,
\begin{eqnarray}
\label{eqn:dipfld}
\hat{\mathcal{E}}_{dip}=-2C \frac{M_i M_j}{|i-j|^3}.
\end{eqnarray}
We have carried out our studies for two representative values of $C$, namely $6 \times 10^{-6}J$ and $2.4 \times 10^{-5}J$ which indirectly correspond to two different intermolecular separations.
\section{Rejection Free Kinetic Monte Carlo Method}
It is well known that the classic Monte Carlo method uses a master equation to obtain equilibrium probabilities of the configurations of a microscopic system at a given temperature \cite{Voter}. Thus the Monte Carlo dynamics cannot be associated with real time and a Monte Carlo step cannot be associated with a time interval and the usual Monte Carlo methods are static methods. However, if we assume that the dynamics in a real system follows Poisson distribution, it is possible to associate a time step with a Monte Carlo step. We obtain the unnormalized cumulative probability for all possible outcomes from a given state and advance time as proportional to the negative log of a uniform random number divided by the cumulative probability. This allows treating the fraction of the cumulative probability for transition to a given state as the probability for the event to occur in a Poisson process. kMC method gives the real dynamics if the process is indeed Poisson and we can scale the time which is in arbitrary units by comparing with a real system with well established model parameters and experimentally known dynamics. In this case, the time will be in actual units. In our case, we do not have such a system for comparison and hence the time will remain in arbitrary units.

We have employed the rejection free kMC method to study the dynamics of magnetization relaxation in the assembly of SMMCs. We have considered $10^5$ SMMCs in the assembly, each SMMC consisting of six spins. The ground state of each SMMC has spin $S=6s$ and $M=+6s$. The states of the Markov chain consist of all the eigenstates of all the SMMCs in the assembly, that is $(2s+1)^6 \times 10^5$ for site spin $s$ in a SMMC. The initial state of the Markov chain has each SMMC in the ground state with $M_i=+6s$. We employ the single spin flip mechanism for accessing various states of the Markov chain. In the implementation of the algorithm, we pick a lattice `$i$' at random (using a uniform random number), the SMMC at that site has energy and magnetization $E_{k,i}$ and $M_{k,i}$, where $k$ labels the eigenstates and for computational convenience is read from a list of the current states $\left|k\right\rangle$ of all the SMMCs in the lattice. We update this list at the end of each kMC step. We choose the possible magnetization of the final state of the chosen SMMC, $M_{l,i}$ to be either $M_{k,i}+1$ or $M_{k,i}-1$, for $-6s<M_{k,i}<6s$ with equal probability. If $M_{k,i}=6s$ [$M_{k,i}=-6s$] we chose $M_{l,i}=6s-1(M_{l,i}=-(6s-1))$ with unit probability. We then select all the states $\left|f\right\rangle$ of the SMMC at site `$i$' with magnetization $M_{l,i}$ and compute the change in energy $\Delta E_{kf}$ for each of these states $\left|f\right\rangle$, correct to first order in perturbation
\begin{eqnarray}
\Delta E_{kf}=\left(E_{f,i}-E_{k,i}\right) + \sum_{j \neq i} B_{dip,j}\left(M_{l,i}-M_{k,i}\right),
\end{eqnarray}
where the summation over all the SMMCs and $B_{dip,j}$ is the local dipolar field given by,
\begin{eqnarray}
B_{dip,j}=-2C \sum_{p\neq j} \frac{M_{m,p}}{|p-j|^3}.
\end{eqnarray}
The quantity $P(f)$ is calculated from $\Delta E_{kf}$ and the temperature of simulation T as, 
\begin{eqnarray}
P(f)=e^{-\Delta E_{kf}/T}.
\end{eqnarray}
We define a cumulative quantity $c(r)$ defined as,
\begin{eqnarray}
c(r)=\sum_{q=1}^r P(q).
\end{eqnarray}
The normalized $\eta(r)$ corresponding to $c(r)$ are given by
\begin{eqnarray}
\eta(r)=\frac{c(r)}{c(L)},
\end{eqnarray}
where `$L$' is the total number of eigenstates with magnetization $M_{l,i}$. We now pick another uniformly distributed random number `$\xi$' and choose the final state of SMMC `$i$' as the state $\left|f\right\rangle$ which satisfies the inequality $\eta(f) < \xi \leq \eta(f+1)$. We employ the binary search scheme for determining the final state, $f$, as it is computationally efficient, particularly for large $L$ when the site spins in SMMC are $3/2$ or $2$. We also update the local dipolar field the end of each MC step for computational efficiency in calculating the change in energy associated with possible final states. At the end of the MC step, we advance the time by $\Delta t$,
\begin{eqnarray}
\Delta t = -\frac{\rm log\zeta}{c(L)}
\end{eqnarray}
where $\zeta$ is another uniformly distributed random number. The kMC evolution is carried out until the magnetization of the assembly is much smaller than $M_0/e$, where $M_0$ is the saturation magnetization given by $6sN$. From the $ln(M(t))$ vs $t$ plot, we can obtain $\tau_r$ the relaxation time at a given temperature and from the plot of $\rm ln(\tau_r)$ vs $1/T$, we estimate the blocking temperature. At low-temperatures, it takes a few billion MC steps for an assembly of $10^5$ SMMCs to significantly relax the magnetization.
\section{Results and Discussion}
We have carried out simulation on assemblies of $10^5$ SMMCs, with each SMMC consisting of six spins, with all spins having the same spin of either $1$, $3/2$ or $2$. We have obtained the relaxation times $\tau_r$ as a function of the exchange anisotropy parameter $\epsilon$, the on-site anisotropy $d$ and the dipolar interaction strength $C$. In the next subsection, we discuss our results for the spin $1$ case and in the subsequent subsection we present the results for the spin $3/2$ and $2$ cases.
\subsection{System with site spins-1}
\begin{figure}
    \includegraphics[width=\columnwidth]{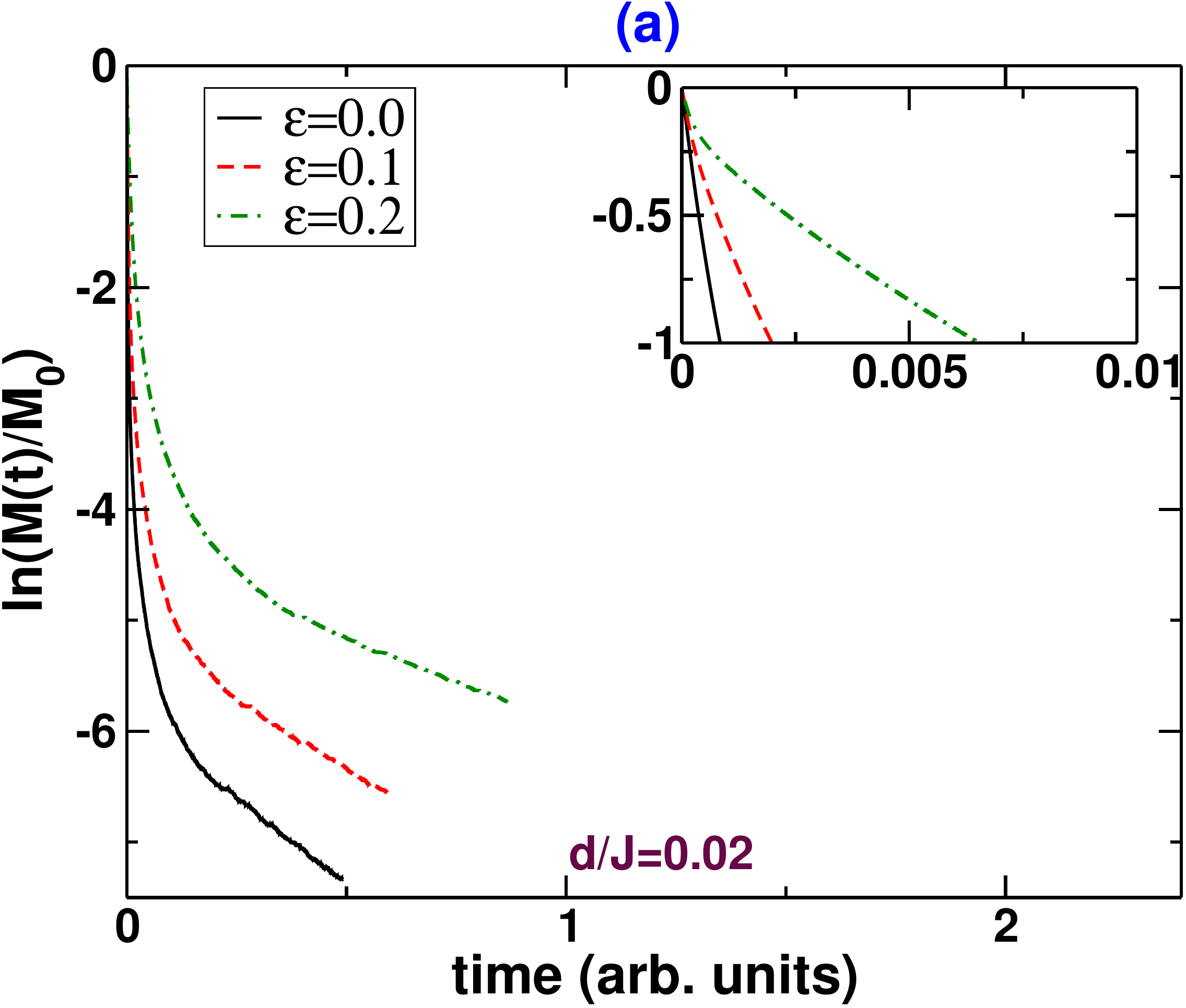}
     \includegraphics[width=\columnwidth]{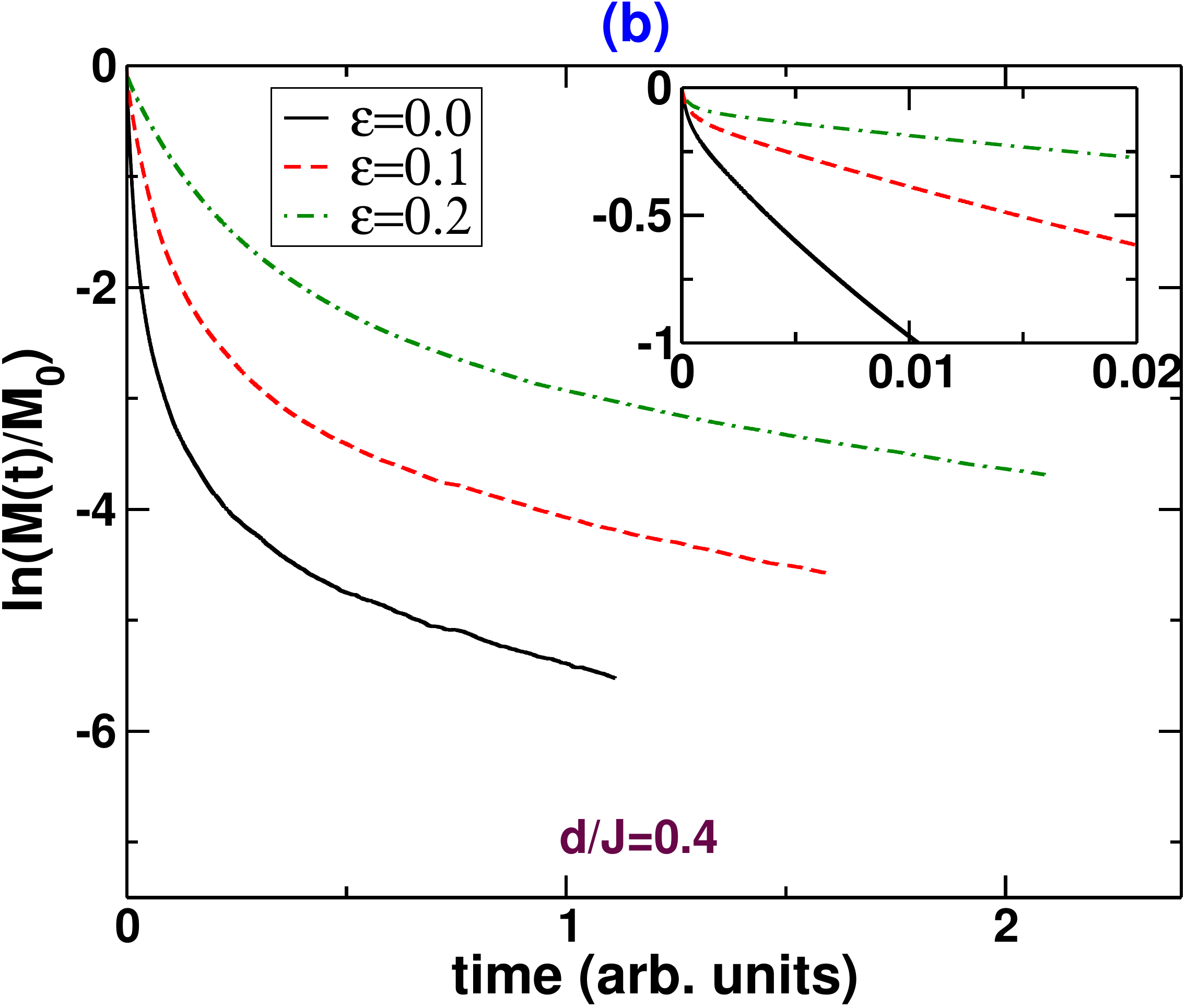}
      \caption{\label{fig:Magvstime}Plot of log magnetization vs time in arbitrary units for an assembly of $10^5$ SMMCs, and site spin $s=1$. Left panel corresponds to $d/J=0.02$ and the right panel to $d/J=0.4$. The spin-dipolar interaction parameter is set at $C=6 \times 10^{-6} J$ and temperature to $0.5\frac{J}{k_B}$ ($J=1$). $M_0$ is the saturation magnetization. Inset in both figures shows the time dependence of initial decay from $M_0$ to $M_0/e$.}
\end{figure}
\begin{figure}
    \includegraphics[width=\columnwidth]{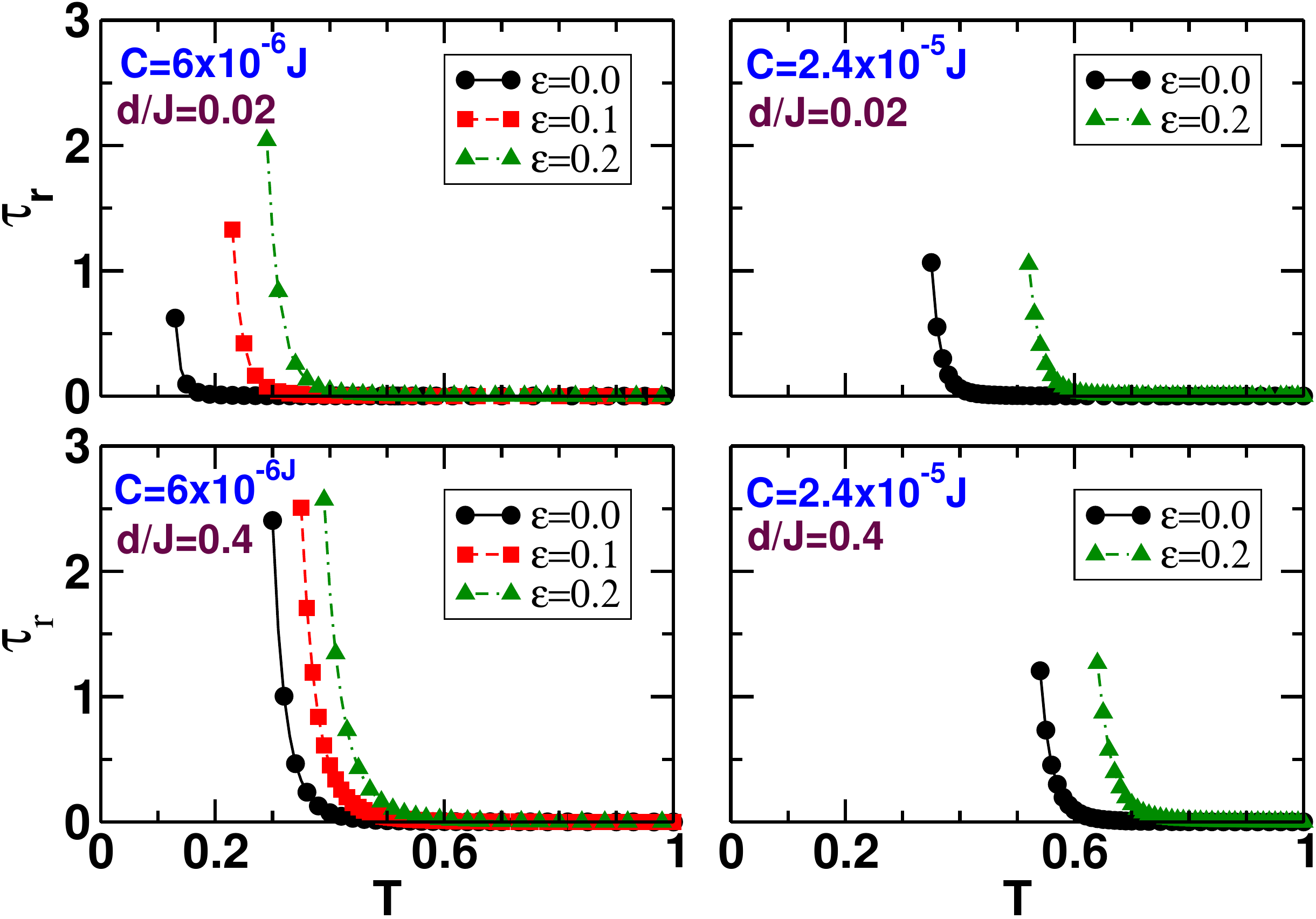}
      \caption{\label{fig:TauvsTemp}Dependence of $\tau_r$ (in arbitrary units) on $d/J$, $C$ and $\epsilon$ for $s=1$ systems. The plot is truncated when $\tau_r$ becomes very large at low temperatures. The temperature is in units of $\frac{J}{k_B}$.}
\end{figure}

In Fig. \ref{fig:Magvstime}, we show the relaxation of magnetization as a function of time for different values of $\epsilon$ for small (left panel) and large on-site anisotropies for dipolar interaction strength of $6 \times 10^{-6}J$. We see that the relaxation occurs very rapidly for small on-site anisotropy almost independent of the strength of exchange anisotropy (Fig. 2a). The relaxation becomes slower for large on-site anisotropy and increasing exchange anisotropy (Fig. 2b). We note that the initial decay in magnetization is exponential and fast. As the time progresses, the decay becomes slower and at long times, the decay is again exponential, but with a much higher relaxation time. We are interested only in the initial decay as within this period the magnetization relaxes to $M_0/e$, where $M_0$ is the saturation magnetization, hence the relaxation time for this decay is the relevant relaxation time. We can obtain the relaxation time $\tau_r$ as a function of temperature and parameters of the model from these plots.
\begin{figure}
    \includegraphics[width=\columnwidth]{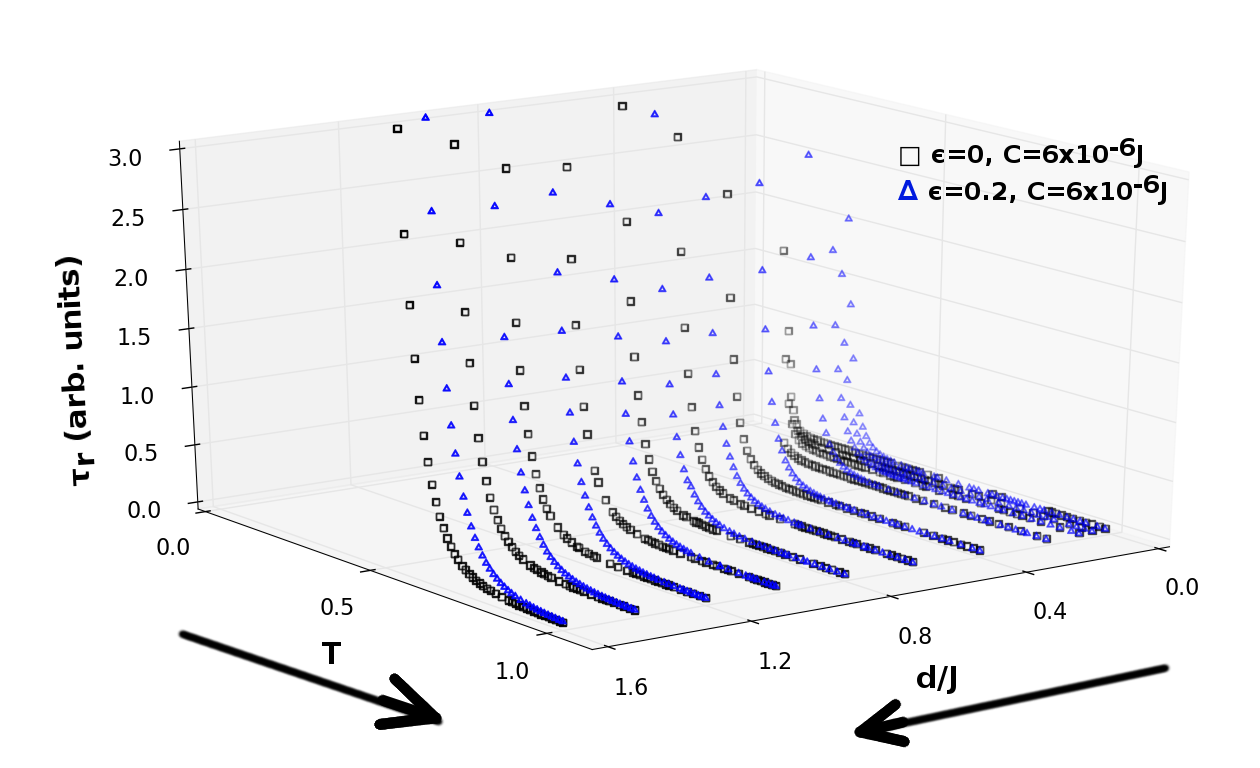}
        \includegraphics[width=\columnwidth]{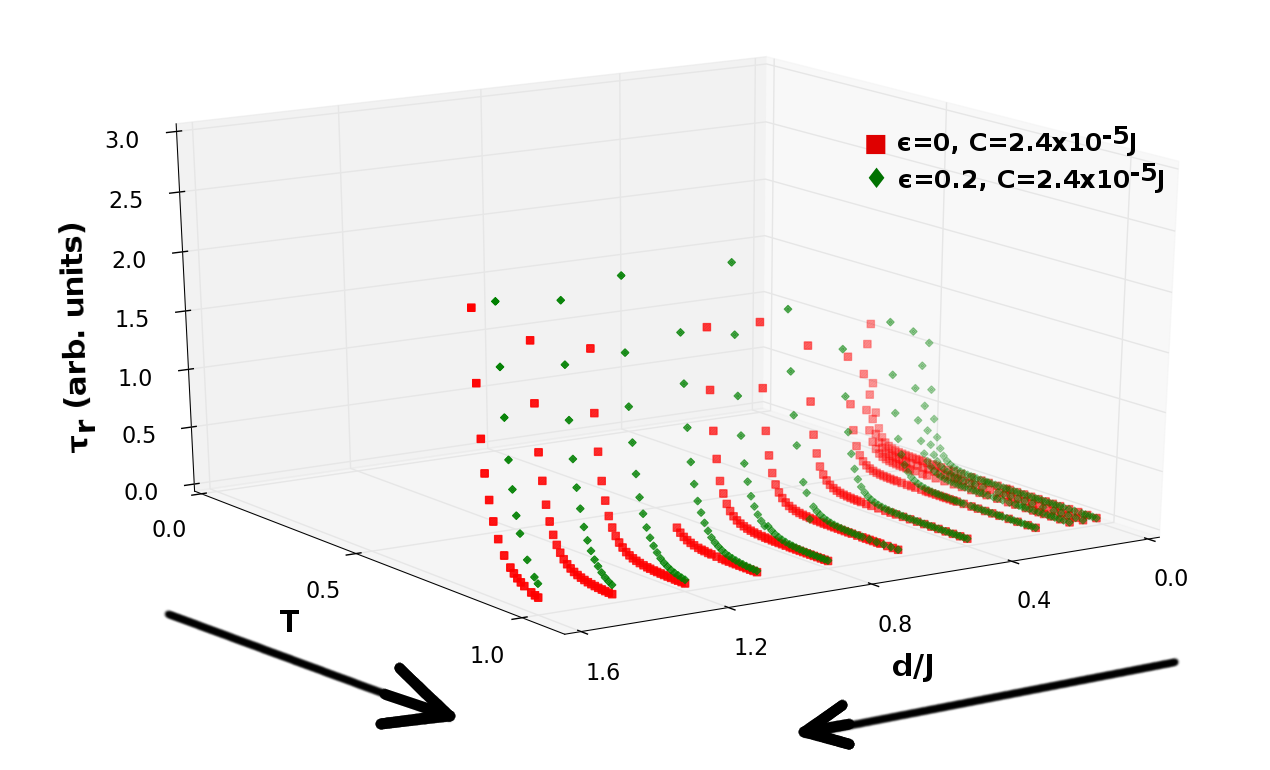}
      \caption{\label{fig:3dplot}3D plot of relaxation time $\tau_r$ (in arbitrary units) of $s=1$ systems vs on-site anisotropy, $d$ and temperature, $T$ (in units of $\frac{J}{k_B}$) for four different values of exchange anisotropy, $\epsilon$ and spin-dipolar interaction strength, $C=6\times 10^{-6}J$ (top) and $C=2.4\times 10^{-5}J$ (bottom).}
\end{figure}

In Fig. \ref{fig:TauvsTemp}, we show the dependence of the relaxation time on the model parameters. The relaxation times increase with increase in on-site anisotropy, $d$, as well as increase in exchange anisotropy, $\epsilon$. However, the dependence of $\tau_r$ on the strength of intermolecular interactions, $C$ is stronger than either on $d$ or on $\epsilon$. The strength of $C$ is dependent on the intermolecular separation as well as on the number of neighbours at any given distance which is determined by the packing arrangement. All the relaxation time results are consolidated in Fig. \ref{fig:3dplot}, where a 3-d plot of $\tau_r$ as a function of temperature and on-site anisotropy for different $C$ and $\epsilon$ values are shown. We see that $\tau_r$ falls off more slowly with temperature as the strengths of interactions go up.
\begin{figure}
    \includegraphics[width=\columnwidth]{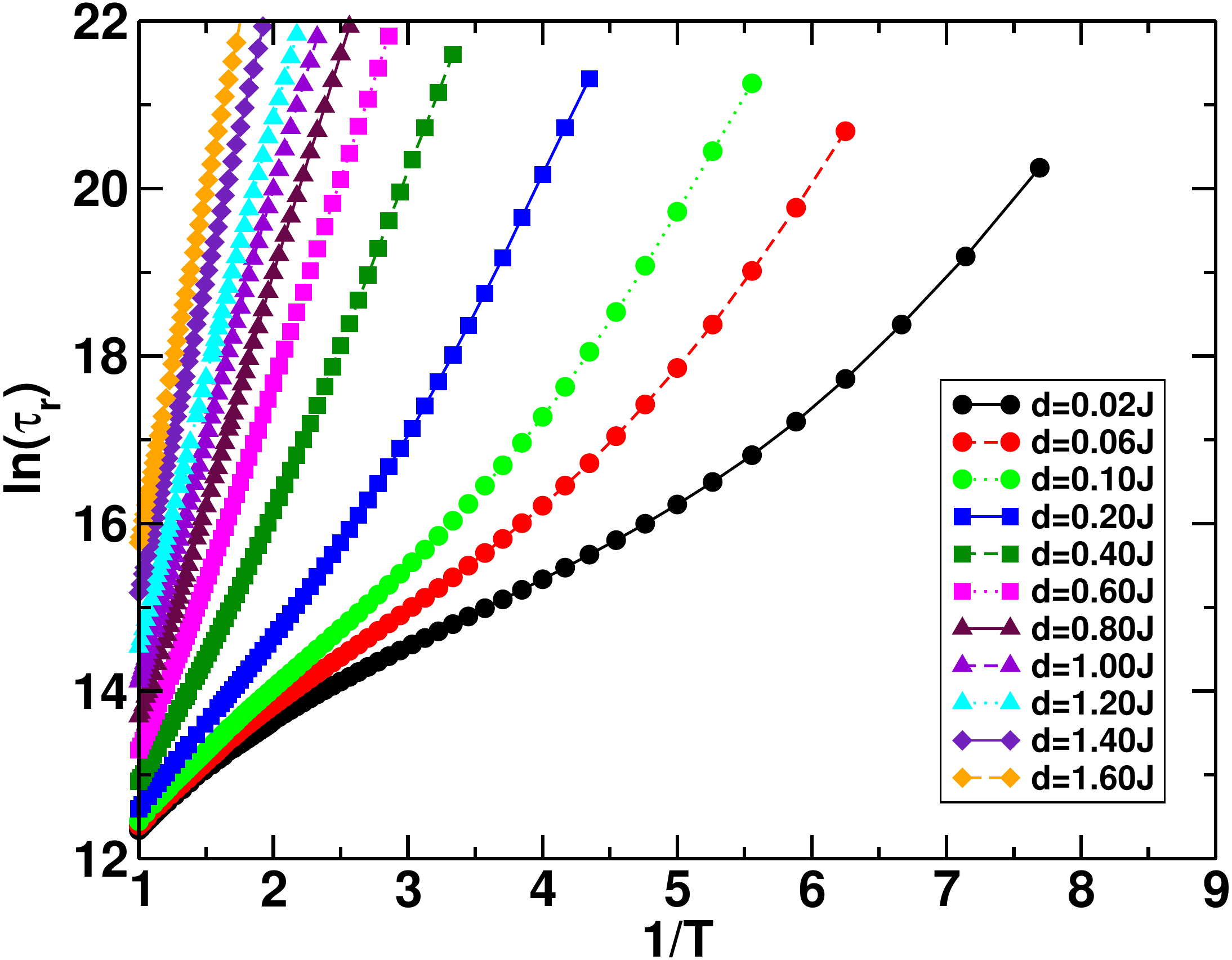}
      \caption{\label{fig:LogtauvsJbyT}Dependence of $\rm ln(\tau_r)$ (in arbitrary units) on $1/T$ (in units of $\frac{k_B}{J}$) for isotropic exchange between spins in a SMMC and inter SMMC interaction parameter is $C=6\times 10^{-6}J$, for different on-site anisotropy strengths.}
\end{figure}
\begin{figure}
    \includegraphics[width=\columnwidth]{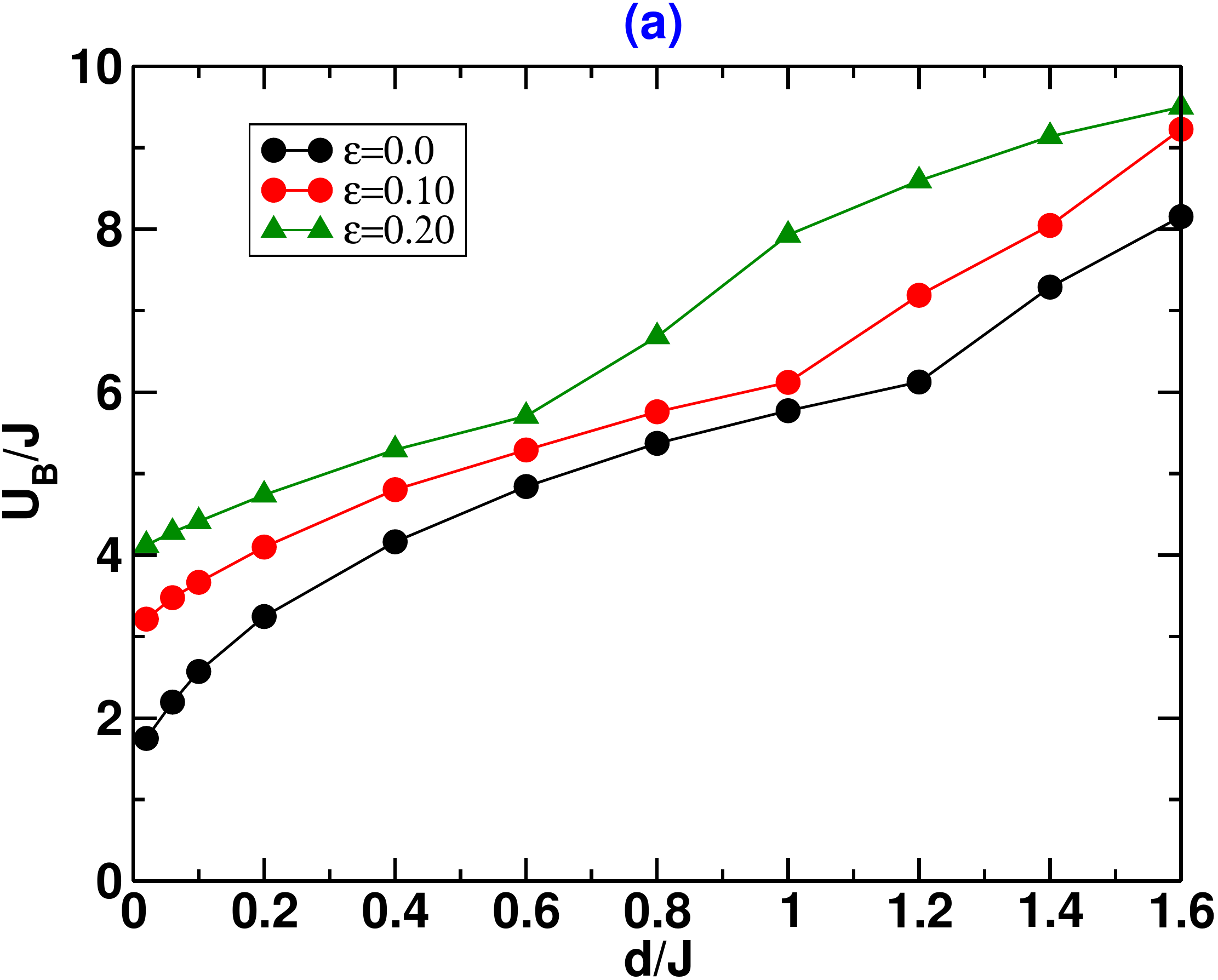}
     \includegraphics[width=\columnwidth]{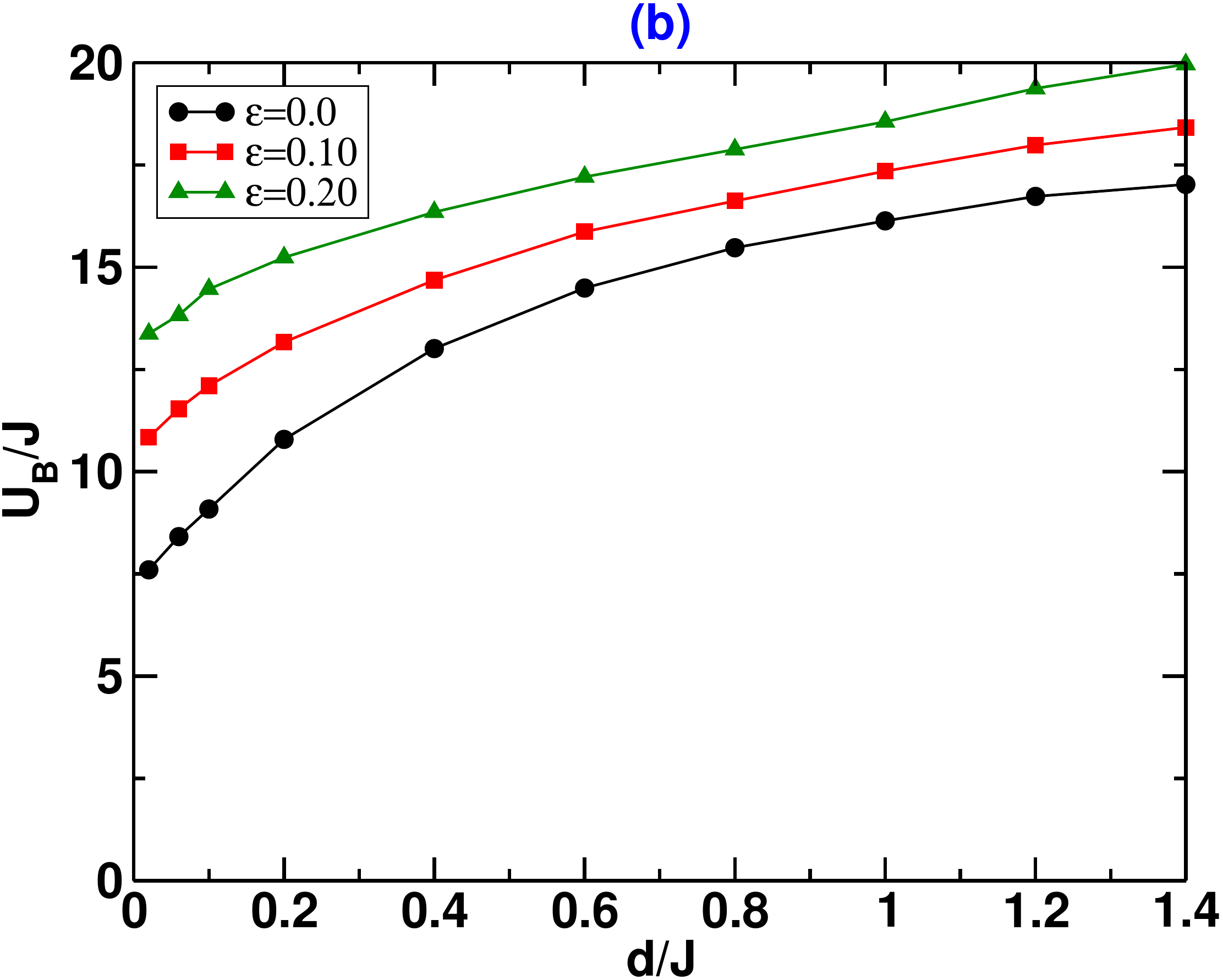}
      \caption{\label{fig:BarriervsdbyJ}Dependence of energy barrier, $U_B$, on $d/J$, the strength of on-site anisotropy, for different exchange anisotropies in the $s=1$ systems. The inter SMMC interaction parameter, $C$, for the left panel is $6\times 10^{-6}J$ and the right panel is $2.4\times 10^{-5}J$.}
\end{figure}

In Fig. \ref{fig:LogtauvsJbyT}, we show the dependence of $\rm ln(\tau_r)$ on $1/T$ for 
isotropic exchange and fixed $C$, the intermolecular spin-spin interaction strength, for various strengths of on-site anisotropy. We note that dependence is nonlinear for on-site anisotropy strength $d/J < 0.6$ \cite{Lampropoulos}. However, for stronger on-site interactions ($d/J > 0.6$) the behaviour is Arrhenius like. Indeed we see similar behaviour even when the exchange interactions are non-isotropic, when intermolecular interactions are stronger. Notwithstanding this nonlinear behaviour, from the low-temperatures we can extract the energy barrier for magnetization relaxation by fitting the data in this region to a straight line, since only the low-temperature behaviour is relevant.

In Fig. \ref{fig:BarriervsdbyJ}, we show the dependence of energy barrier as a function of on-site anisotropy and exchange anisotropy. We note that in both cases, the energy barrier tends to saturate for large on-site anisotropies. In the isotropic exchange model, there is slightly more rapid increase in the energy barrier to relaxation with increase in on-site anisotropy. However, this dependence becomes weaker as the exchange anisotropy is increased. We find that as the inter SMMC spin dipolar interaction strength is quadrupled there is roughly a three-fold increase in the energy barrier for small $d/J$. However, at large $d/J$ this increase is only two-fold. This goes to show that increase in spin-dipolar interaction strength reduces the dependence of the energy barrier on the on-site anisotropy parameter $d/J$.
\subsection{System with site spins 3/2 and 2}
\begin{figure}
    \includegraphics[width=\columnwidth]{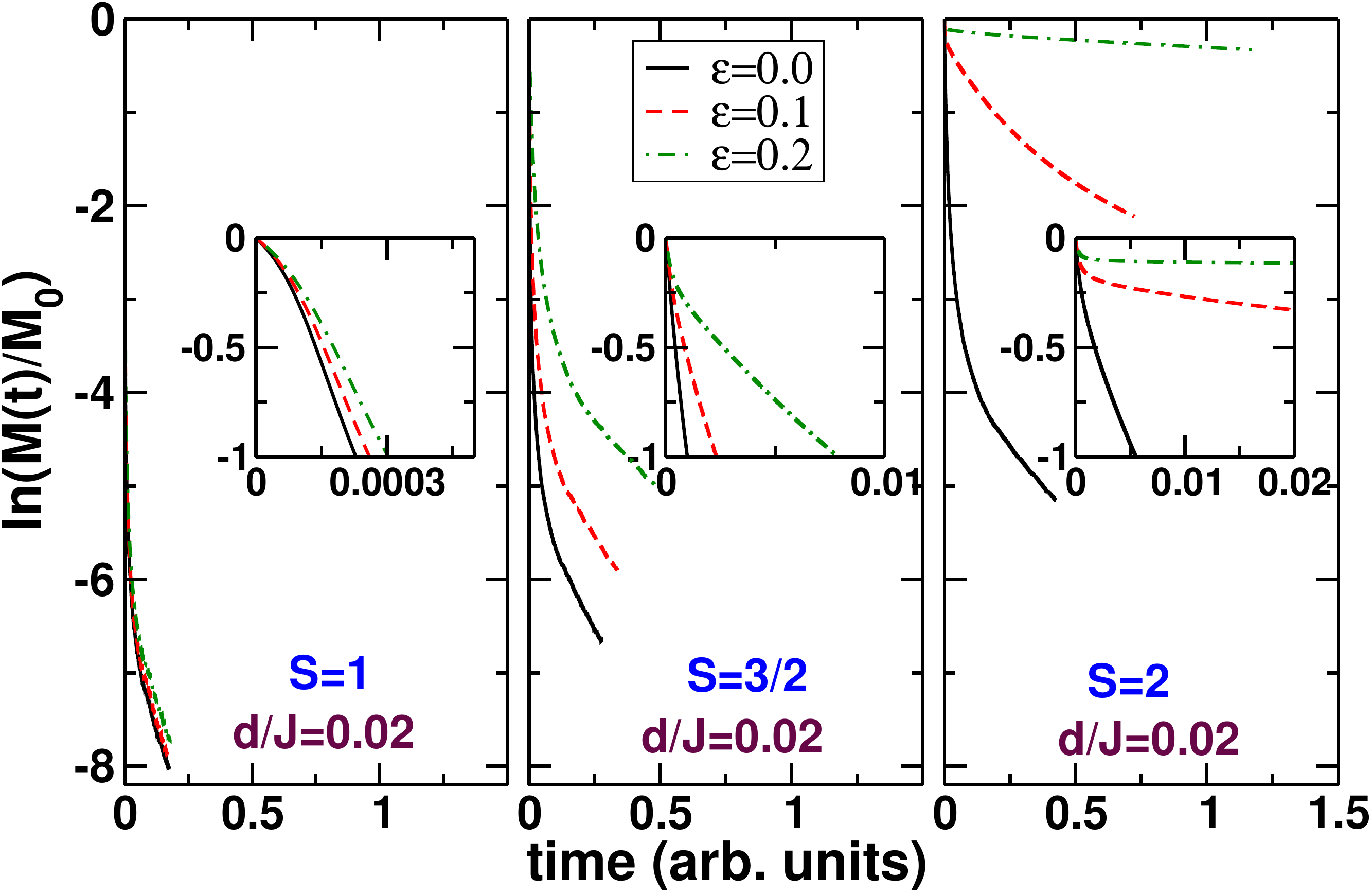}
          \caption{\label{fig:MagvstimediffSpin}Log of magnetization relaxation vs time for $d/J=0.02$, at $T=1$ (in units of $J/k_B$), and dipolar interaction parameter $C=6\times 1^{-6}J$ for three different anisotropic exchange values. We see that on the same scale, $s=1$ system relaxes extremely fast as can be seen from the inset.}
\end{figure}
\begin{figure}
    \includegraphics[width=\columnwidth]{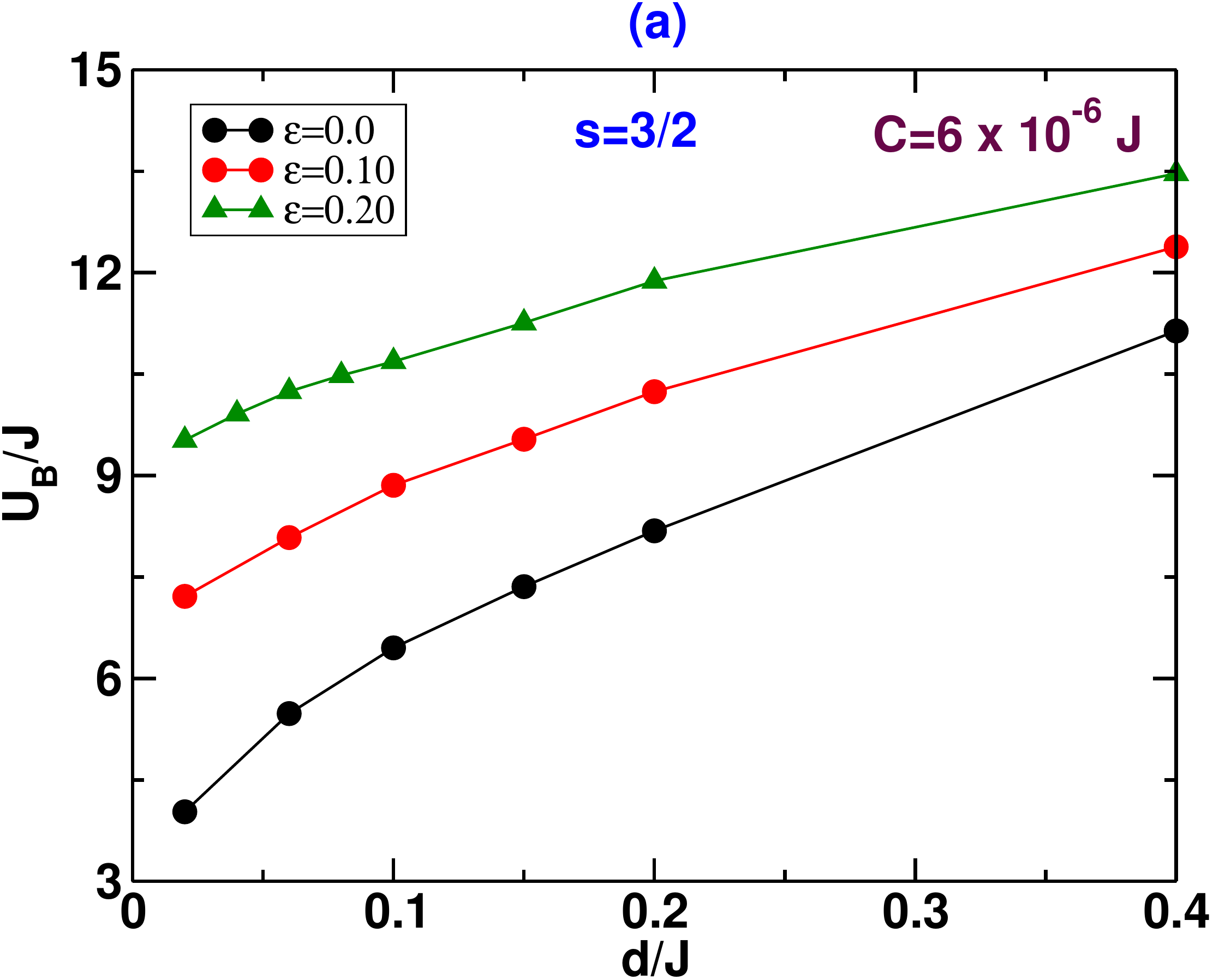}
     \includegraphics[width=\columnwidth]{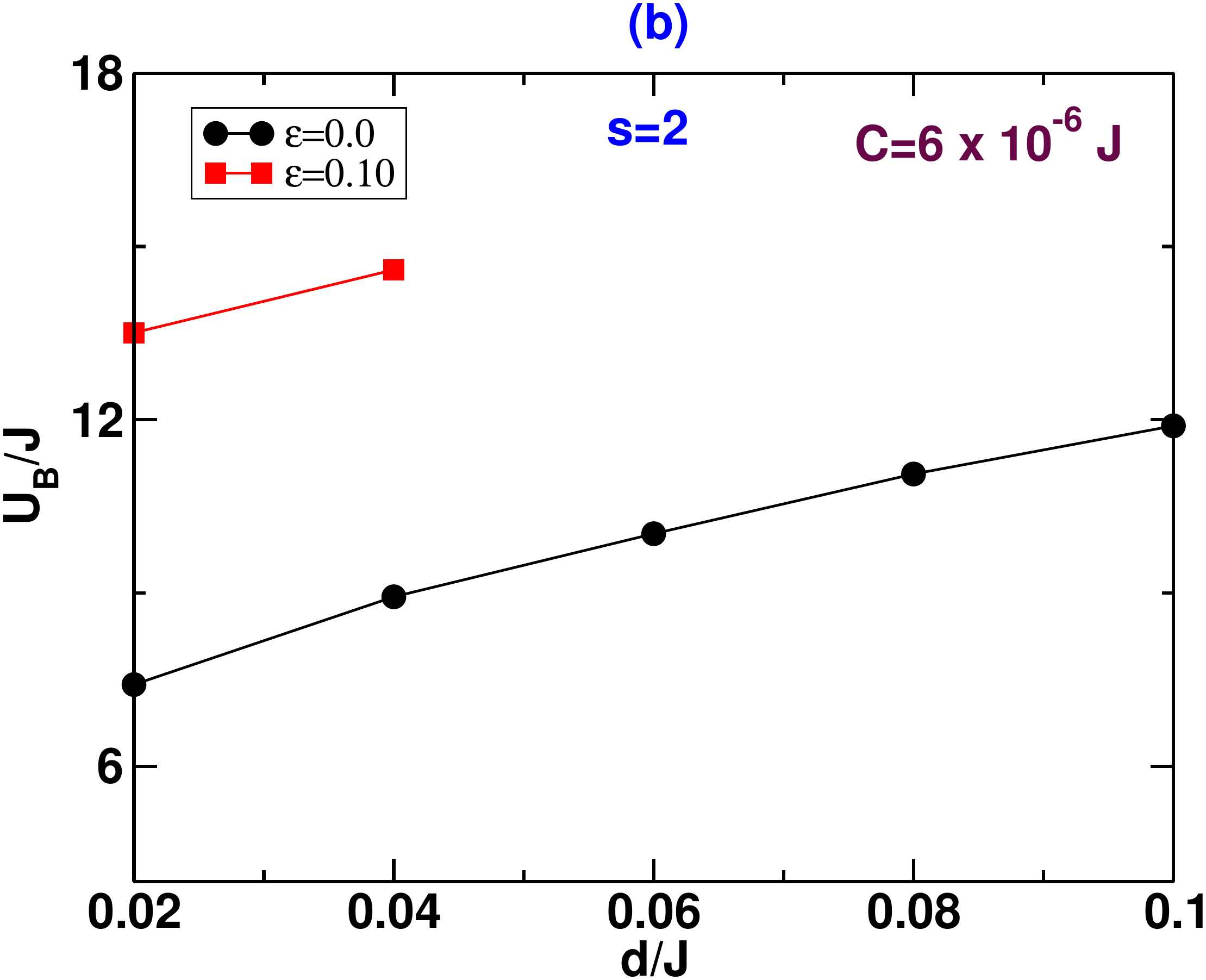}
      \caption{\label{fig:BarriervsdbyJdiffSpin}Dependence of energy barrier, $U_B$ on on-site anisotropy for an assembly of $10^5$ SMMC on a chain. Left panel is for $s=3/2$ and the right panel is for $s=2$; the spin-dipolar interaction strength is fixed at $C=6\times 10^{-6}J$. In the $s=3/2$ case, we could compute the barrier for the three exchange anisotropy values and $d/J$ up to 0.4. However, in the $s=2$ case, we could compute the barrier only up to $d/J=0.1$ for the isotropic exchange case and up to $d/J=0.04$ for anisotropic exchange with $\epsilon=0.1$. For higher exchange anisotropy and large on-site anisotropy the magnetization did not relax in a reasonable computational time.}
\end{figure}
In Fig. \ref{fig:MagvstimediffSpin}, we have shown the magnetization for small on-site anisotropy, for three different exchange anisotropies for all the three spin systems, $s=1$, $3/2$ and $2$. We note that the exchange anisotropy hardly influences the speed of relaxation in the $s=1$ case, while in higher spin systems there is a strong dependence of the relaxation time on anisotropy. In fact, in the $s=2$ system for large exchange anisotropy, the relaxation is too slow to obtain a relaxation time with the computer resources available to us. Indeed, we could relax the magnetization within a reasonable computational time, only for a few cases in the $s=3/2$ and $s=2$ systems. In Fig. \ref{fig:BarriervsdbyJdiffSpin}, we have shown the dependence of the energy barrier on  $d$ for several exchange anisotropies for the $s=3/2$ and $s=2$ cases. Indeed the magnetization also does not relax in reasonable computational time for the larger strength of intermolecular interactions, namely $C=2.4\times 10^{-5}J$. While in the regime of small on-site anisotropy, the barrier quickly saturates, we note a strong dependence of relaxation time on the exchange anisotropy. Thus, clearly the barrier height depends upon the on-site anisotropy, $d$, exchange anisotropy, $\epsilon$, site spin and strength of spin-dipolar interactions, $C$, but the dependence on $C$ is stronger than on either $\epsilon$ or $d$. This is because the dipolar interaction energy scales as $S^2$ leading to increase in the energy barrier to relaxation.  

Our studies show that to design a single chain magnet with high barrier to magnetization relaxation, we need to have as high a spin of the SMMC as possible. Besides, we should have reasonably large on-site anisotropy and large exchange anisotropy. Importantly, we need a large spin-dipolar interaction strength, which in turn implies tight packing of SMMC and a high site spin. The highest known energy barrier to magnetic relaxation is found in single-ion rare earth magnetic molecules. These systems being rare earth ion systems have both high on-site anisotropy and high spin in the ground state. Since the magnetic molecules contain only one rare earth ion, they are relatively small molecules and the packing tends to be closer and tighter packing results in stronger dipolar interactions. All these factors favour a large thermal barrier to magnetization relaxation in these systems. 
\section{Summary and Conclusions}
We have carried out an innovative rejection free kMC simulation to study the dependence of the barrier to magnetization relaxation on on-site anisotropy, exchange anisotropy and spin-dipolar interactions. The model system consists of $10^5$ SMMCs each with anisotropic exchange interactions between uniaxially anisotropic site spins of magnitude $1$, $3/2$ and $2$. The SMMCs experience spin dipolar interactions. We have used all the model exact eigenstates of all the individual SMMCs in an assembly of $10^5$ of them arranged on a one-dimensional lattice to carry out kMC simulations within a single spin-flip mechanism. The SMMCs interact with each other via a spin-dipolar interaction and are arranged so as to yield a ferromagnetic ground state. We relax the ferromagnetic ground state at different temperatures using the rejection free kMC algorithm. We obtain the magnetization relaxation time as a function of temperature at different points in the parameter space of the model. We find the energy barrier saturates with increase in on-site uniaxial anisotropy, in every case. The barrier is larger for larger exchange anisotropy, higher site spin and larger strength of spin-dipolar interactions. The magnetization does not relax appreciably for higher spins even for small on-site anisotropy. However, the energy barrier, where it could be computed, saturates rapidly with on-site anisotropy. The energy barrier to relaxation also increases with exchange anisotropy and has a strong dependence on the strength of spin-dipolar interactions. This is because both the spin dipolar interaction energy and the barrier height of an isolated SMMC scale as $S^2$ where $S$ is the total spin of the SMMC at a lattice site. For ferromagnetic exchange interactions the spin of the low-lying states of a SMMC scales linearly with site spin $s$. As a result, even after several billion MC steps magnetization does not relax appreciably from saturated magnetization at temperatures, in SMMCs with higher site spin and large uniaxial anisotropy. We attribute the large energy barrier in the recently discovered rare earth single ion magnets is due to large spin-dipolar interactions arising from small size of the molecule as well as due to large single ion anisotropy and high spin in the ground state.

This study has focused on individual SMMCs chains arranged on a 1-d lattice. We need to extend these studies to real molecules such as $\textrm Mn_{12}Ac$, $\textrm Fe_{8}$. We are also engaged in extending these studies to 2-d and 3-d packings to identify the lattice feature that lead to large energy barriers to thermal relaxation of magnetization.
\section{Acknowledgement}
SR thanks DST, India for financial support and INSA for a fellowship. SH thanks DST for a fellowship which supported this work.

\end{document}